\documentclass[12pt]{article}                  
\usepackage{amsfonts, amsmath, amssymb, amsthm, amsbsy}
\theoremstyle{remark}
\newtheorem{rem}{Remark}
\newtheorem*{exampleNo}{Example}

\newcommand{\cE}{\mathcal{E}}

\newcommand{\cR}{\mathcal{R}}

\newcommand{\vph}{\varphi}

\newcommand{\dd}{\partial}
\newcommand{\sd}{\dd_\theta}
\newcommand{\oh}{\tfrac{1}{2}}

\newcommand{\by}[1]{\textit{{#1}}}
\newcommand{\jour}[1]{\textit{{#1}}}
\newcommand{\vol}[1]{\textbf{{#1}}}
\newcommand{\book}[1]{\textrm{{#1}}}

\begin{document}

\title
{On weakly non\/-\/local, nilpotent, and super\/-\/recursion
operators for $N=1$ super\/-\/equations}

\date{November 25, 2005}

\author{A.\,V.\,Kiselev%
\thanks{\textit{Permanent address}:
Department of Higher Mathematics,\ Ivanovo State
Power University,\ Rabfakovskaya str.\,34, Ivanovo, 153003
Russia.
\textit{Current address}:
Department of Physics, Middle East Technical University,
06531 Ankara, Turkey.
E-mail: \texttt{arthemy\symbol{"40}newton.physics.metu.edu.tr}.
}%
${\ }^{,}$\thanks{%
Proc.\ Int.\ Workshop ``Supersymmetries and Quantum Symmetries--05,''
JINR, Dubna, 
2005.},
\ T. Wolf%
\thanks{
Department of Mathematics, Brock University, 500 Glenridge Ave.,
St.~Catharines, Ontario, Canada L2S~3A1.
E-mail: \texttt{twolf\symbol{"40}brocku.ca}.
}
}
\maketitle

\begin{abstract}
\noindent%
We consider nonlinear, scaling\/-\/invariant $N=1$ boson$+$fermion
supersymmetric systems whose right\/-\/hand sides are homogeneous
differential polynomials and satisfy some natural assumptions.
We select the super\/-\/systems that admit infinitely many higher
symmetries generated by recursion operators; we further restrict
ourselves to the case when the dilaton dimensions of the
bosonic and fermionic
super\/-\/fields coincide and the weight of the time is half
the weight of the spatial variable.
We discover five systems that satisfy these assumptions; one system is
tran\-s\-for\-med to the purely bosonic Burgers equation. We construct
local, nilpotent, triangular,
weakly non\/-\/local, and
super\/-\/recursion operators for their symmetry algebras.

\noindent%
2000 MSC:
  35Q53, 
  37K05, 
  81T40. 

\noindent%
\textit{Key words and phrases}: Supersymmetric recursion operators,
Burgers equation.
\end{abstract}

\paragraph*{Introduction.}
We consider the problem of a complete description of $N=1$ nonlinear,
scaling invariant
evolutionary super\/-\/equations $\{f_t=\phi^f$, $b_t=\phi^b\}$
that admit infinitely many symmetries
$\{f_s=F$, $b_s=B\}$ proliferated by recursion
operators~$\cR$; here $b$ is the bosonic super\/-\/field and $f$ is
the fermionic super\/-\/field.
The axioms for selecting $N=1$ nonlinear homogeneous polynomial
evolutionary systems with higher symmetries
were suggested~\cite{SUSY} by V.\,V.\,Sokolov and A.\,S.\,Sorin;
the axioms are discussed in~\cite{Kiev2005}.

By construction, the equations
are scaling invariant: their right\/-\/hand sides are
differential polynomials homogeneous w.r.t.\ a set of (half-)integer
weights $[\theta]\equiv-\oh$, $[x]\equiv-1$,
$[t]<0$, $[f]$, $[b]>0$;
we also assume that the negative weight $[s]$ is (half\/-\/)integer.
Here we denote by $\theta$ the super\/-\/variable
and we put $\sd\equiv D_\theta+\theta\,D_x$
such that $\sd^2=D_x$; here $D_\theta$ and $D_x$ are the total
derivatives w.r.t.\ $\theta$ and $x$, respectively.
All notions and notation follow~\cite{JKKersten},
see also~\cite{Kiev2005} for details.

In this paper, we investigate the properties of systems such that the
weight of the 
time $t$ is $[t]=-\oh$.\ We also assume
$[f]=[b]=\oh$ (the weights may not be uniquely defined).

The first version of \textsc{SsTools} package~\cite{WolfCrack} for
\textsc{Reduce} was used for finding the systems that satisfy the above
axioms and possess higher symmetries under the bound
$-5\leqslant[s]\leqslant-\oh$. Five systems were thus discovered, see
Table~I below. Later, we used the second version of
\textsc{SsTools}~\cite{CompPhysCommun} for symmetry analysis of the
super\/-\/systems in~\cite{SUSY} and for constructing conservation laws
and recursion operators for their symmetry algebras.
The method of Cartan forms~\cite{JKKersten} for the recursion
operators was applied.
Within this approach, the recursions are regarded as symmetries of the
linearized equations. Namely, we `forget' the internal structure of the
symmetry flows $f_s=F$, $b_s=B$ and operate with $F$ and $B$ as
we do with the
components of solutions of the linearized equations.
The expressions $\cR=R(F,B)$ are the recursion operators if
each $\cR$
satisfies the linearized equation again and if they are linear
w.r.t.\ $F$, $B$, and their derivatives.

Let us introduce some notation.
Assume $\cR$ is a recursion for an equation and consider the symbol
${}_{\text{ord}}^{\text{layers}}\cR_{\text{weight}}^{\sharp}$.
The subscripts `ord' and `weight' denote the differential order and the
weight of the recursion $\cR$, respectively,
and the superscript `layers' (if
non\/-\/empty) indicates the required
number of layers of the nonlocal variables
assigned to conservation laws. The symbol `$\sharp$' denotes the number
of recursions for a given differential order,
weight, and the nonlocalities.
Further, we denote by $L$ the local recursion operators,
by $N$ the nonlocal or weakly non\/-\/local \cite{TMPhGallipoli,
Novikov} recursions, the symbol $Z$ denotes a nilpotent
recursion whose powers equal zero except for a finite set, and
$\varSigma$ is a super\/-\/recursion that swaps the parities of the
flows.

Now we list the five new
super\/-\/equations and indicate their recursions.
The weights of the recursion operators are calculated w.r.t.\ the
standard values $[f]=[b]=\oh$, $[t]=-\oh$.

\centerline{
\begin{tabular}{|
  l | 
  c |}
\hline
\eqref{BurgSystem}\quad
$\left\{ \begin{aligned}f_t&=\sd{b},\\ b_t&=b^2+\sd{f} \end{aligned}
\right.$ &   ${}_1^1 N_{-1}^1$ \\
\hline
\eqref{DoubleLayer}\quad
$\left\{\begin{aligned}f_t&=\sd{b}+fb,\\ b_t&=\sd{f}\end{aligned}
\right.$ &
${}_0^2 N_{-1\frac{1}{2}}^1$,\ ${}_{\frac{1}{2}}^2 N_{-2}^1$,\
${}_{\frac{1}{2}}^2 N_{-2\frac{1}{2}}^1$,\ ${}_{\frac{1}{2}}^2 N_{-3}^1$ \\
\hline
\eqref{Quad}\quad
$\left\{\begin{aligned}f_t&=-\alpha\,fb,\\ b_t&=b^2+\sd{f}\end{aligned}
\right.$ &
$\begin{aligned}
\alpha&=2:\quad
 {}_{\frac{1}{2}} L_{-2}^{1},\
{}_{\frac{1}{2}} L_{-2\frac{1}{2}}^1,\ {}_{0}^0 Z_{-2}^1,\ 
{}_{0}^0 \varSigma_{-2}^1,\ {}_{0}^0 \varSigma_{-2\frac{1}{2}}^1,\ %
{}_{0}^0 \varSigma_{-2\frac{1}{2}}^1;\phantom{\Bigl(} \\
\alpha&=1:\quad{}
{{}_{0}^0 Z_{-2\frac{1}{2}}^1,\ {}_{\frac{1}{2}} Z_{-3}^1};
\qquad{\alpha=4:\quad
{}_{1} L_{-3\frac{1}{2}}^1}\phantom{\Bigl(}
\end{aligned}$
\\
\hline
\end{tabular}%
}

\smallskip\centerline{Table~I.}

\smallskip
It turns out that these equations
exhibit practically the whole variety of properties that superPDE of
mathematical physics possess. 
Let us discuss the properties of the equations present in Table~I in
more detail.


\paragraph*{1. The Burgers equation.}
First we construct an $N=1$ supersymmetric representation of
the Burgers equation and investigate its properties.
We consider the system 
\begin{equation}\label{BurgSystem}
f_t=\sd{b},\qquad b_t=b^2+\sd{f}.
\end{equation}
There is a unique set of weights
$[f]=[b]=\oh$, $[t]=-\oh$, $[x]=-1$ in this case.
Hence we conclude that the above system precedes
the invariance w.r.t.\ the translation along~$x$.
Equation~\eqref{BurgSystem} admits the continuous
sequence~\eqref{SymBurg} of higher
symmetries $f_s=\phi^f$, $b_s=\phi^b$ at all
(half-)integer weights $[s]\leq-\oh$.    
Also, there is the continuous sequence~\eqref{BurgSSym}
of supersymmetries for
Eq.~\eqref{BurgSystem} at all (half-)integer weights
$[\bar{s}]\leq-\oh$ of the fermionic `time'~$\bar{s}$.

System~\eqref{BurgSystem} is obviously reduced to the purely bosonic
Burgers equation
$b_x=b_{tt}-2bb_t$.
We emphasize that the role of the independent coordinates $x$ and $t$
is reversed w.r.t.\ the standard interpretation of $t$ as
the time and $x$ as the spatial variable.
The Cole\/--\/Hopf substitution $b=-u^{-1}u_t$ from the heat equation
$u_x=u_{tt}$
is thus the solution for the bosonic component of~\eqref{BurgSystem}.

Further, we introduce the bosonic nonlocality $w$ of weight $[w]=0$
by the rules $\sd{w}=-f$, $w_t=-b$.
The variable $w$ is a potential for both fields $f$ and $b$.
The nonlocality satisfies the potential Burgers equation
$w_x=w_{tt}+w_t^2$ such that the formula $w=\ln u$ gives the solution;
the relation $f=-\sd{w}$ determines the fermionic component in
system~\eqref{BurgSystem}.

Now we extend the set of dependent variables $f$, $b$, and $w$ by the
symmetry generators
$F$, $B$, and $W$ that satisfy the linearized relations
upon the flows of the initial super\/-\/fields, respectively.
In this setting, we obtain the recursion
\begin{equation}\label{RecBurgers}
\cR_{[1]}=\binom{F_x-\sd{f}\,F+f_x\,W}%
                        {B_x-\sd{f}\,B+b_x\,W}\ \Longleftrightarrow\ %
R=\begin{pmatrix}
D_x-\sd{f}+f_x\,\dd_\theta^{-1} & 0 \\
b_x\,\dd_\theta^{-1} & D_x-\sd{f}
\end{pmatrix}
\end{equation}
of weight $[s_R]=-1$.
In agreement with~\cite{TMPhGallipoli}, the above recursion is
{weakly non\/-\/local}~\cite{Novikov}.
We recall that a recursion operator $R$ is \emph{weakly non\/-\/local}
if each nonlocality $\partial_\theta^{-1}$ is preceded with a (shadow
\cite{JKKersten} of a nonlocal) symmetry $\vph_\alpha$ and is followed
by the gradient $\psi_\alpha$ of a conservation law:
$R=\text{local
part}+\sum_\alpha\vph_\alpha\cdot\partial_\theta^{-1}\circ\psi_\alpha$.
From \cite{TMPhGallipoli} it follows that this property is
automatically satisfied by all recursion operators which are
constructed using one layer of the nonlocal variables assigned to
conservation laws.

Recursion~\eqref{RecBurgers} generates two sequences of higher
symmetries for system~\eqref{BurgSystem}:
\begin{equation}\label{SymBurg}
\binom{f_t}{b_t}\mapsto
\binom{\sd{b_x}-\sd{f}\sd{b}-f_xb}{\sd{f_x}-(\sd{f})^2-b^2\sd{f}+bb_x}
  \mapsto\cdots,\ 
\binom{f_x}{b_x}\mapsto
\binom{f_{xx}-2\sd{f}f_x}{b_{xx}-2\sd{f}b_x} \mapsto \cdots.
\end{equation}
Also, recursion~\eqref{RecBurgers} produces two infinite sequences
of supersymmetries for~\eqref{BurgSystem}:
\begin{equation}\label{BurgSSym}
\binom{\sd{f}}{\sd{b}} \mapsto
\binom{\sd{f_x}-(\sd{f})^2-f_xf}{\sd{b_x}-\sd{f}\,\sd{b}-b_xf}
\mapsto\cdots, \qquad
\binom{f\sd{b}-b\,\sd{f}+b_x}{b\sd{b}-f\,\sd{f}+f_x-fb^2}
\mapsto\cdots.
%
\end{equation}

\begin{rem}
System~\eqref{BurgSystem} is not a supersymmetric extension of
the Burgers equation;   
it is a supersymmetric representation of
the Burgers equation.
However, symmetries~\eqref{SymBurg} and~\eqref{BurgSSym}
are \emph{not} reduced to the
purely bosonic $(x,t)$-independent symmetries~\cite{Lychagin}
of the Burgers equation (particularly, owing to
the interchanged role of the variables $x$ and~$t$).
We finally recall that the Burgers equation  
has infinitely many higher symmetries that depend explicitly on the base
coordinates $x$, $t$ but exceed the set of axioms~\cite{Kiev2005}
we use.
\end{rem}

Two supersymmetric generalizations ($N=0$ and $N=2$)
of the Burgers equation are constructed
in~\cite{Kiev2005}. The $N=0$ extension relates it with integrable
flows on associative algebras. The $N=2$ Burgers equation contains a
KdV\/-\/type component and admits an $N=2$ modified KdV equation as a
symmetry flow.

\paragraph*{2. A system with nonlocal recursions.}
The second system, 
\begin{equation}\label{DoubleLayer}
f_t=\sd{b}+fb,\qquad    b_t=\sd{f},
\end{equation}
is also homogeneous w.r.t.\ a unique set of weights
$[f]=[b]=\oh$, $[t]=-\oh$, $[x]=-1$.
Similarly to the supersymmetric representation~\eqref{BurgSystem}
for the Burgers equation, 
Eq.~\eqref{DoubleLayer} admits
symmetries $(f_s$, $b_s)$ for all weights~$[s]\leq-\oh$. 

We conjecture that system~\eqref{DoubleLayer}
has only one conservation law that
defines the fermionic variable~$w$ of weight~$0$ by
$ 
w_t=f$, $\sd{w}=b$.
Then, many nonlocal conservation laws and hence many new
variables appear. We use the fermionic variable~$v$ whose
weight $[v]=\tfrac{3}{2}$ is minimal: we set
$v_t=\sd{b}\cdot wfb+f_xwf$ and
$\sd{v}=-\sd{b}\cdot fb+\sd{f}\cdot\sd{b}\cdot w+b_xwf$.
Now, there are nontrivial solutions to the determining equations
for recursion operators. First, we obtain the recursion of
zero differential order with nonlocal coefficients: 
\[
R_{[-1\oh]} =
\binom{-\sd{b}\cdot wf B + wvF  + v\cdot B}%
      {\sd{b}\, wf F  - vF + vw\cdot B}.
\]
Also, we get a nonlocal operator with nonlocal coefficients, 
\[
R_{[-2]} =
\binom{\sd{b}\,Vw-\sd{f}\,\sd{B}\,wf-\sd{f}\,\sd{b}\,Wf+\sd{f}\,\sd{b}\,Fw
    +\sd{f}\,V+Vwfb}%
 {\sd{B}\,\sd{b}\,wf+\sd{b}\,V-\sd{b}\,Fwfb+\sd{f}\,\sd{b}\,Vwf+\sd{f}\,Vw-Vfb}.
\]
The coefficients of the recursions found for
$[s_R]=-2\oh$ and $[s_R]=-3$ are also nonlocal.


\paragraph*{3. A triplet of super\/-\/systems.}
Finally, we consider the three systems
\begin{equation}\label{Quad}
f_t=-\alpha fb,\qquad b_t=b^2+\sd{f}
\end{equation}
which differ by the values $\alpha=1$, $2$, and $4$ of the
parameter~$\alpha$ and therefore exhibit rather different properties.
The weights for the above equation are multiply defined, and we
choose the tuple $[f]=[b]=\oh$, $[t]=-\oh$, $[x]=-1$ to be the primary
`reference system.'

\subparagraph*{Case $\alpha=2$.}
First, we fix $\alpha=2$ 
and consider Eq.~\eqref{Quad}: we get
$ 
f_t=-2fb$, $b_t=b^2+\sd{f}$.
The weights for symmetries are $[s]=-\oh$, $[s]=-1$, and then
Eq.~\eqref{Quad} admits a continuous chain of symmetry flows
for all (half-)integer weights $[s]\leq-2\oh$.  
Surprisingly, no nonlocalities are needed to construct the recursion
operators, although there are many conservation laws for this system.
We obtain purely local recursion operators~$\cR$ that
proliferate the symmetries:
$\vph=(F,B)\mapsto\vph'=\cR$ for any $\vph$.
The recursion          
\[
\cR_{[-2]} =
\begin{pmatrix}
\tfrac{11}{2}\sd{F}\,\sd{f}\,f + 11\sd{F}\,fb^2 + \tfrac{3}{2}(\sd{f})^2 F +
3\sd{f}\,Fb^2 + \tfrac{1}{2}f_xFf \\
  \begin{gathered}[t]  {} \\
   11\sd{B}\,fb^2 + 8\sd{b}\,Fb^2 + 22\sd{b}\,fBb + 7(\sd{f})^2 B +{}\\
   14\sd{f}\,Bb^2 + \tfrac{11}{2}\sd{f}\,\sd{B}\,f +
   \tfrac{5}{2}\sd{f}\,\sd{b}\,F +
   \tfrac{1}{2}b_xFf + f_x F b + 5 f_xfB
  \end{gathered}
\end{pmatrix},
\]
of weight $[s_R]=-2$ is triangular since $R^f$ does not contain~$B$.
Also, we obtain the recursion of weight~$2\oh$; its components are
\begin{align*}  
\cR_{[-2\oh]}^f &=
  - 2\sd{b}\,Ffb^2 - \sd{F}\,\sd{f}\,fb - \sd{F}\,fb^3 - \tfrac{1}{2}f_xFfb
  - 2\sd{f}\,fBb^2,\\
\smash{\cR_{[-2\oh]}^b} &=
  \sd{B}\,fb^3 + \sd{b}\,Fb^3 + \sd{b}\,fBb^2 + \tfrac{1}{8}\sd{f_x}Ff +{}\\
 &\quad+ \tfrac{1}{2}\sd{F}b^4 + \tfrac{1}{2}\sd{F}(\sd{f})^2
  + \sd{F}\,\sd{f}\,b^2 + \tfrac{1}{8}\sd{F}\,f_xf + (\sd{f})^2Bb +{}\\
 &\quad+ \sd{f}\,Bb^3 + \sd{f}\,\sd{B}\,fb + \sd{f}\,\sd{b}\,Fb
  + \sd{f}\,\sd{b}\,fB + \tfrac{3}{8}\sd{f}\,F_xf + {}\\
 &\quad+\tfrac{1}{4}\sd{f}\,f_xF
  + \tfrac{1}{2}b_xFfb + \tfrac{1}{2}F_xfb^2 + \tfrac{1}{4}f_xFb^2
  + \tfrac{1}{2}f_xfBb.
\end{align*}
Further, we get
a triangular nilpotent operator of weight $-3$ such that 
$\cR_{[-3]}^f=0$ and $\cR_{[-3]}^b=
{(\sd{f})^3fF+6(\sd{f})^2fb^2F+12\sd{f}\,fb^4F+8fb^6F}$.
The above recursion is a recurrence relation~\cite{Kiev2005}
which is well\/-\/defined for all symmetries of Eq.~\eqref{Quad}.
Another local
recursion for $[s]=-3$ is huge and therefore omitted. 


For $\alpha=2$, system~(\ref{Quad})
admits at least three super\/-\/recursions
${}^t(R^f$, $R^b)$ such that the parities of $R^f$ and
$R^b$ are opposite to the odd parity for $f$ (and hence for $F$) and to
the even parity of $b$ and~$B$. This property
is possible owing to the presence of the \emph{odd} variable~$s_R$.
The triangular zero\/-\/order super\/-\/recursions are 
${\bar{\cR}}_{[-2]}^f =   
{4\sd{f}\,Ffb+8Ffb^3}$, ${\bar{\cR}}_{[-2]}^b=
{-4\sd{b}\,Ffb+2(\sd{f})^2F+6\sd{f}\,Fb^2+4\sd{f}\,fBb-f_xFf+4Fb^4+8fBb^3}$
and   
\[
{\bar{\cR}}_{[-2\oh]} =
\binom{-\sd{f}\,f_xF-2f_xFb^2}%
{\sd{b}\,f_xF-\sd{f}\,b_xF+\sd{f}\,f_xB-2b_xFb^2+2f_xBb^2}
\]
for weights $[s_R]=-2$ and $[s_R]=-2\oh$, respectively;
the third super\/-\/recursion found
for $[s_R]=-2\oh$ is very large.       
Quite naturally, system~\eqref{Quad} has infinitely many
supersymmetries if~$\alpha=2$.

\subparagraph*{Case $\alpha=1$.}
For $\alpha=1$ from~\eqref{Quad} we obtain the system 
$ 
f_t=-fb$, $b_t=b^2+\sd{f}$.
The default set of weights is the same as above:
$[f]=[b]=\oh$, $[t]=-\oh$, and $[x]=-1$.
The sequence of symmetries is not continuous
and starts later than for the chain in the case $\alpha=2$.
We find out that there are symmetry flows
if either $[s]=[t]=-\oh$ (the equation itself),
$[s]=[x]=-1$ (the translation along~$x$),
or $[s]\leq-3\oh$ such that a
continuous chain starts for all (half-)\/in\-te\-ger
weights~$[s]$. 

Similarly to the previous case, no nonlocalities are needed to
construct the recursions, which therefore are purely local.
The recursion operator   
$\cR_{[-2\oh]}^f=0$, $\cR_{[-2\oh]}^b=
{(\sd{f})^2\,Ff+3\sd{f}\,Ffb^2+\tfrac{9}{4}Ffb^4}$
of maximal weight $[s_R]=-2\oh$ is nilpotent: $\cR^2=0$.
For the succeeding weight $[s_R]=-3$,
we obtain a nilpotent local recursion with components
\begin{align*} 
\cR_{[-3]}^f&=
  \tfrac{5}{3}\sd{F}\,(\sd{f})^2f+\tfrac{5}{2}\sd{F}\,\sd{f}\,fb^2
  -\tfrac{5}{3}(\sd{f})^3F-\tfrac{5}{2}(\sd{f})^2Fb^2 +{}\\
 &+{5}\sd{f}\,\sd{b}\,Ffb+\tfrac{20}{3}\sd{f}\,f_xFf
  +\tfrac{15}{2}f_xFfb^2,\\
\cR_{[-3]}^b&=
  \sd{f_x}\,Ffb-\tfrac{105}{2}\sd{F}\,\sd{b}\,fb^2
  -\tfrac{160}{3}\sd{F}\,\sd{f}\,\sd{b}\,f+{11}\sd{F}\,f_xfb+{}\\
 &+\tfrac{5}{3}(\sd{f})^2\sd{B}f+\tfrac{5}{3}(\sd{f})^2\sd{b}\,F
  +\tfrac{5}{2}\sd{f}\,\sd{B}\,fb^2+\tfrac{5}{2}\sd{f}\,\sd{b}\,Fb^2-{}\\
 &-{55}\sd{f}\,\sd{b}\,fBb+\tfrac{17}{3}\sd{f}\,b_xFf
  +\sd{f}\,f_xfB+\tfrac{23}{2}b_xFfb^2+\tfrac{183}{2}f_xfBb^2.
\end{align*}
It generates symmetries of system~(\ref{Quad});
the differential order of $\cR_{[-3]}$ is positive.

\subparagraph*{Case $\alpha=4$.}
Finally, let $\alpha=4$; then system~\eqref{Quad}
acquires the form   
$ 
f_t=-4fb$, $b_t=b^2+\sd{f}.$
Again, the basic set of weights is
$[f]=[b]=\oh$, $[t]=-\oh$, $[x]=-1$, and system~\eqref{Quad}
admits the symmetries $(f_s$, $b_s)$ such that their weights are
$[s]=-\oh$, $-1$ or $[s]\leq-3\oh$ w.r.t.\ the basic set.  
This situation coincides with the case~$\alpha=1$.
Again, no nonlocalities are needed for constructing the recursion
of minimal weight $[s_R]=-3\oh$:
\begin{align*} 
\smash{\cR_{[-3\oh]}^f}&=
   -12\sd{b}\,Ffb^4-\sd{F}\,(\sd{f})^2fb
   -{4}\sd{F}\,\sd{f}\,fb^3 - 3\sd{F}\,fb^5-{}\\
  &-{4}(\sd{f})^2fBb^2-{4}\sd{f}\,\sd{b}\,Ffb^2
   -\tfrac{2}{3}\sd{f}\,f_xFfb-12\sd{f}\,fBb^4-{2}f_xFfb^3,\\
\smash{\cR_{[-3\oh]}^b}&=
   3\sd{B}fb^5+3\sd{b}\,Fb^5+9\sd{b}\,fBb^4
   +\tfrac{1}{9}\sd{f_x}\,\sd{f}\,Ff-\tfrac{1}{3}\sd{f_x}\,Ffb^2+\tfrac{3}{4}\sd{F}\,b^6+{}\\
  &+\sd{F}\,\sd{b}\,fb^3
   +\tfrac{1}{4}\sd{F}\,(\sd{f})^3+\tfrac{5}{4}\sd{F}\,(\sd{f})^2b^2
   +\tfrac{7}{4}\sd{F}\,\sd{f}\,b^4+\sd{F}\,\sd{f}\,\sd{b}\,fb+{}\\
  &+\tfrac{5}{18}\sd{F}\,\sd{f}\,f_xf
   +\tfrac{1}{2}\sd{F}\,f_xfb^2+(\sd{f})^3Bb
   +{4}(\sd{f})^2Bb^3+(\sd{f})^2\sd{B}\,fb+{}\\
  &+(\sd{f})^2\sd{b}\,Fb+(\sd{f})^2\sd{b}\,fB
   +\tfrac{2}{9}(\sd{f})^2F_xf+\tfrac{1}{6}(\sd{f})^2f_xF+3\sd{f}\,Bb^5+{}\\
  &+{4}\sd{f}\,\sd{B}\,fb^3
   +{4}\sd{f}\,\sd{b}\,Fb^3 + {10}\sd{f}\,\sd{b}\,fBb^2+\tfrac{2}{3}\sd{f}\,b_xFfb
   +\sd{f}\,F_xfb^2+{}\\
  &+\tfrac{2}{3}\sd{f}\,f_xFb^2
   +\tfrac{5}{3}\sd{f}\,f_xfBb
   +{2}b_xFfb^3+F_xfb^4+\tfrac{1}{2}f_xFb^4+f_xfBb^3.
\end{align*}
No nilpotent recursion operators were found for system~(\ref{Quad})
if~$\alpha=4$.

\begin{rem}\label{InfByNilpotent}
We discovered that an essential part of recursion operators for
supersymmetric PDE are nilpotent.
At present, it is not clear how the nilpotent recursion operators
contribute to the integrability of supersymmetric systems and what
invariants they describe or symptomize.
Further, we emphasize that this property does not always originate from
the rule `$f\cdot f=0$', but this is an immanent feature of the
symmetry algebras.
More generally, the nilpotent recursions are quite natural in the
bosonic sector, too. We have
\end{rem}

\begin{exampleNo}[I. S. Krasil'shchik, private communication]
Consider a system of linear ordinary differential equations
$\dot{\boldsymbol{x}}=A(t)\,\boldsymbol{x}$. Then any nilpotent
constant matrix $R$ that commutes with the matrix~$A$ is a recursion.
\end{exampleNo}

It would be of interest to construct an equation~$\cE$ that admits
nilpotent differential recursion operators $\{R_1,\ldots\mid
R_i^{n_i}=0\}$ which generate an infinite sequence of symmetries
$\vph$, $R_{i_1}(\vph)$, $R_{i_2}\circ R_{i_1}(\vph)$, $\ldots$
for~$\cE$. Here we assume that
at least two operators (without loss of generality, $R_1$ and $R_2$)
do not commute and hence the flows never become zero.

\paragraph*{Acknowledgements}
The authors thank I.\,S.\,Krasil'shchik, A.\,S.\,Sorin,
and A.\,M.\,Ver\-bo\-vet\-sky for stimulating discussions.

\end{document}